\def\rfr#1{eq. (\ref{#1})}

\def\bar{\begin{eqnarray}}
\def\ear{\end{eqnarray}}

\def\eqi{\begin{equation}}
\def\eqf{\end{equation}}
\def\eqia{\begin{eqnarray}}
\def\eqfa{\end{eqnarray}}
\def\rp#1#2{{#1\over#2}}

\def\lb#1{\label{#1}}

\def\oc2{$\mathcal{O}(c^{-2})$}

\documentclass[11pt]{article}
\usepackage{amsmath,amsthm,amscd,amssymb}
\usepackage{latexsym}
\usepackage{graphicx,epsfig}

\begin{document}

\noindent{\bf \LARGE{Evidence of the gravitomagnetic field of Mars
}}
\\
\\
\\
{Lorenzo Iorio}\\
{\it Viale Unit$\grave{a}$ di Italia 68, 70125\\Bari, Italy
\\tel./fax 0039 080 5443144
\\e-mail: lorenzo.iorio@libero.it}

\begin{abstract}
According to general relativity, the gravitomagnetic
Lense-Thirring force of Mars would secularly shift the orbital
plane of the Mars Global Surveyor (MGS) spacecraft  by an amount
of 1.5 m, on average,  in the cross-track direction over 5 years.
The determined cross-track post-fit residuals of MGS, built up by
neglecting just the gravitomagnetic force in the dynamical force
models and without fitting any empirical cross-track acceleration
which could remove the relativistic signal, amount to 1.6 m, on
average, over a 5-years time interval spanning from 10 February
2000 to 14 January 2005. The discrepancy with the predictions of
general relativity is 6$\%$.
\end{abstract}

\\
\section{Introduction}
In this communication we propose to explain  certain measured
features of the path of the Martian orbiter Mars Global Surveyor
(MGS) (Konopliv et al. 2006), whose orbital parameters are quoted
in Table \ref{tavola}, with the action of the general relativistic
gravitomagnetic force on the spacecraft. If confirmed by
subsequent, dedicated analyses it would be, in general, the first
test of general relativity in the gravitational field of a Solar
System body other than the Sun and the Earth and, in particular,
the first independent test of the Lense-Thirring effect with
respect to those conducted so far with the LAGEOS satellites in
the Earth's gravitational field (Iorio and Morea 2004, Ciufolini
and Pavlis 2004, Iorio 2006).
\section{The Lense-Thirring effect}
The Lense-Thirring effect consists of tiny secular precessions of
the longitude of the ascending node $\Omega$ and the argument of
pericentre $\omega$ of a test particle freely orbiting a central
rotating body. They are (Lense and Thirring 1918; Barker and
O'Connell 1974; Soffel 1989; Ashby and Allison 1993; Iorio 2001)
\begin{equation}
\left\{
\begin{array}{lll}
\dot\Omega_{\rm LT}&=&\rp{2GS}{c^2 a^3 (1-e^2)^{3/2}},\lb{leti}\\\\
\dot\omega_{\rm LT}&=&-\rp{6GS\cos i}{c^2 a^3 (1-e^2)^{3/2}},
\end{array}
\right.
\end{equation}
where $G$ is the Newtonian gravitational constant, $S$ is the spin
angular momentum of the central body which acts as source of the
gravitational field, $c$ is the speed of light in vacuum, $a, e$
and $i$ are the semimajor axis, the eccentricity and the
inclination of the orbital plane to the central body's equator,
respectively. In general, the gravitmagnetic force affects the
in-plane along-track and the out-of-plane cross-track parts of the
orbit of a the test particle. Indeed, the shifts along the radial
$R$, along-track $T$ and cross-track $N$ directions can be, in
general, expressed in terms of the Keplerian orbital elements as
(Christodoulidis et al. 1988)
\begin{equation}
\left\{
\begin{array}{lll}
\Delta R &=&\sqrt{ (\Delta a)^2 + \rp{[ (e\Delta a + a\Delta e)^2 + (ae\Delta{\mathcal{M}})^2 ]}{2}},\\\\
\Delta T &=& a\sqrt{ 1+\rp{e^2}{2} }\left[\Delta{\mathcal{M}}
+\Delta\omega+\cos i\Delta\Omega +\sqrt{(\Delta
e)^2+(e\Delta{\mathcal{M}})^2} \right],\\\\
\Delta N &=& a\sqrt{ \left(1+\rp{e^2}{2}\right)\left[\rp{(\Delta i
)^2}{2}+(\sin i\Delta\Omega)^2\right] },
%,
\end{array}
\right.
\end{equation}
where $\mathcal{M}$ is the mean anomaly. The Lense-Thirring shifts
are, thus
\begin{equation}
\left\{
\begin{array}{lll}
\Delta R_{\rm LT} &=& 0,\\\\
\Delta T_{\rm LT} &=& a\sqrt{ 1+\rp{e^2}{2}
}\left(\Delta\omega_{\rm LT}+\cos i\Delta\Omega_{\rm LT}\right),\\\\
\Delta N_{\rm LT} &=& a\sqrt{1+\rp{e^2}{2}}\sin i\Delta\Omega_{\rm
LT}.
%,
\end{array}
\right.
\end{equation}
For a near polar satellite like MGS only the out-of-plane shift
$\Delta N_{\rm LT}$ is present.
\section{The interpretation  of the MGS data}
In (Konopliv et al. 2006) six years of MGS Doppler and range
tracking data and three years of Mars Odyssey Doppler and range
tracking data were analyzed in order to obtain information about
several features of the Mars gravity field summarized in the
global solution MGS95J.

\begin{figure}
\begin{center}
\includegraphics[width=13cm,height=11cm]{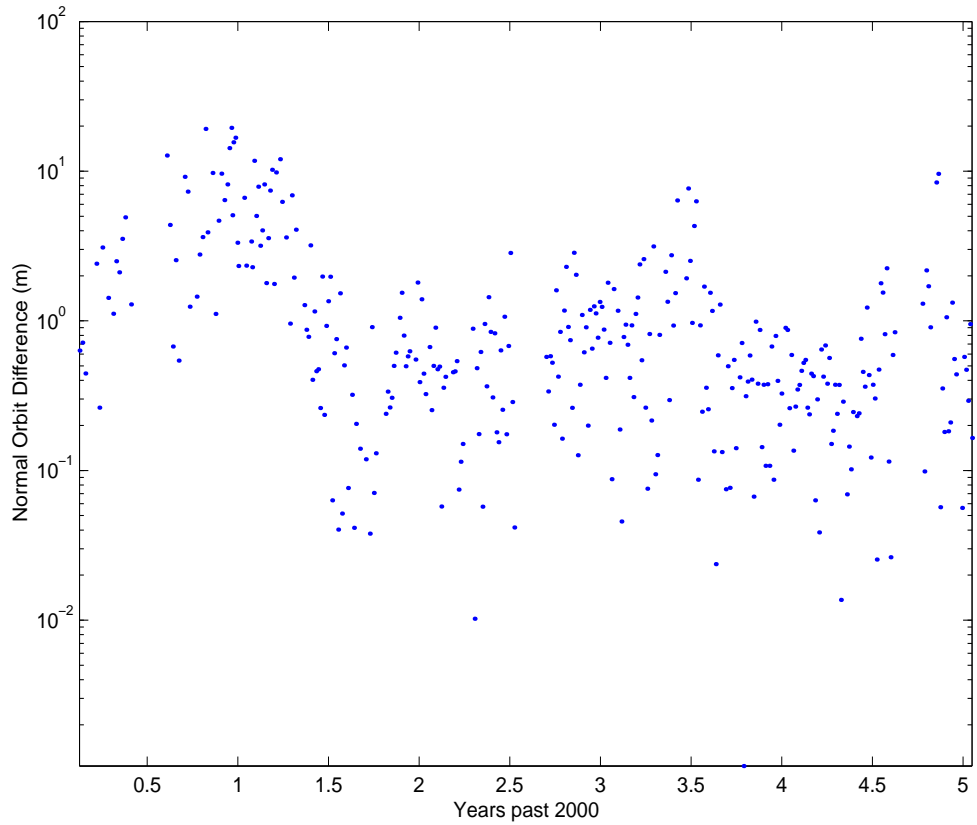}
\end{center}
\caption{\label{plot} Orbit overlap differences (10 February
2000-14 January 2005) for the out-of-plane direction (Data
provided by A. Konopliv, 2006). The average amounts to 1.6 m: the
average Lense-Thirring out-of-plane shift, over the same time
span, is 1.5 m.}
\end{figure}

As a by-product of such an analysis, also the orbit of MGS was
determined with great accuracy (Konopliv et al. 2006). Here we are
interested in particular in the out-of-plane portion of its orbit.
{\small\begin{table}\caption{Orbital parameters of MGS and its
Lense-Thirring precession. $a$ is the semimajor axis, $i$ is the
inclination, and $e$ is the eccentricity. $\dot\Omega_{\rm LT}$ is
the gravitomagnetic node precession and $\left<\Delta N_{\rm
LT}\right>$ is the average out-of-plane gravitomagnetic shift over
five years. }\label{tavola}
\begin{center}
\begin{tabular}{ccccc}
%{\hrule height 1.5pt}
\hline
$a$ & $i$ & $e$ & $\dot\Omega_{\rm LT}$ & $\left<\Delta
N_{\rm LT }\right>$ \\
\hline $3792.42$ (km) & $92.86$ (deg)& $0.0085$ & 33.8 (mas
yr$^{-1}$)&
$1.5$ (m)\\
\hline
%
%
%{\hrule height 1.5pt}
\end{tabular}
\end{center}
\end{table}}
The  root-sum-square  out-of-plane residuals over a five-years
time interval spanning from 10 February 2000 to 14 January 2005
are shown in Figure \ref{plot}.
They have been determined in a Mars-centered coordinate system
defined by the inertial International Celestial Reference Frame
(ICRF), which is nearly equivalent to the Earth's mean equator at
the epoch J2000, by using all the previously estimated parameters
of the global solution. Due to improved modelling (orientation,
gravity, angular momentum wheel de-saturations, and atmospheric
drag), the average of such residuals amounts to 1.6 m
%and it is
%possible to visually discern a secular trend with a slope of
%roughly 0.8 m yr$^{-1}$ or less.

This result can be well explained with the action of the
gravitomagnetic field of Mars on the orbit of MGS. Indeed, by
using for the spin of Mars the value\footnote{It comes from
$S=\alpha MR^2 (2\pi/T)$, where the polar moment is
$\alpha=C/MR^2=0.3654$ (Konopliv et al 2006) and the sidereal
rotation period is $T=1.02595675$ d (Seidelmann 1992).}
$S=1.9\times 10^{32}$ kg m$^2$ s$^{-1}$, the Lense-Thirring
precession of the node of MGS amounts to\footnote{The formulas of
\rfr{leti} hold in a frame whose reference plane is the central
body's equator. This is just the case because the Mars equator is
tilted only by 1.7 deg to the Earth's equator, which is the
reference plane of the frame used in the analysis of (Konopliv et
al. 2006). Indeed, the obliquity of Mars and the Earth amount to
25.2 deg and 23.5 deg respectively. } $\dot\Omega_{\rm LT}=33.8$
milliarcseconds per year (mas yr$^{-1}$), which yields a shift in
the out-of-plane direction $\Delta N_{\rm LT}=a\sin
i\Delta\Omega_{\rm LT}=0.6$ m yr$^{-1}$. This means that over five
years the orbital plane of MGS was shifted by 3 m by the
gravitomagnetic field of Mars, a quantity which falls within the
reached orbit accuracy. The averaged Lense-Thirring shift over
five years is thus 1.5 m.

To corroborate this interpretation of the data, it is important to
note that in the suite of dynamical force models used to reduce
the gathered observations general relativity is, in fact, present,
but only with the gravitoelectric Schwarzschild part: the
gravitomagnetic Lense-Thirring part is not included, so that it is
fully accounted for by the showed residuals. Moreover, no
empirical out-of-plane accelerations were fitted in the data
reduction process; this was done only for the along-track
component of MGS orbit to account for other non-gravitational
forces like thermal, solar radiation pressure and drag
mismodelling.
%But the gravitomagnetic force leaves the along-track
%component of the MGS orbit almost unaffected because of its polar
%geometry.
Thus, it is unlikely that the out-of-plane Lense-Thirring signal
was removed from the data in the least-square procedure. Moreover,
if the relativistic signature was removed so that the determined
out-of-plane residuals were only (or mainly) due to other causes
like mismodelling or unmodelling in the non-gravitational forces,
it is difficult to understand why the along-track residuals
(middle panel of Figure 3 of (Konopliv et al. 2006)) have almost
the same magnitude, since the along-track component of the MGS
orbit is much more affected by the non-gravitational accelerations
(e.g. the atmospheric drag) than the out-of-plane one.
%The even
%zonal harmonic coefficients $J_{\ell}$ of the multipolar expansion
%of the classical part of the Martian gravitational potential
%induce secular node precessions which have the same signature of
%the Lense-Thirring effect and are much larger. However, they are
%just one of the main goals of the MGS mission; since they have
%been solved for their action has been removed from the
%
\section{Conclusions}
In this communication we showed that the general relativistic
gravitomagnetic Lense-Thirring effect can accommodate 94$\%$ of
the 1.6 m average error in the determined out-of-plane orbit
residuals of the Martian orbiter MGS.
 The collection and the processing of
more data from MGS, Odyssey and the recently launched MRO
spacecraft will be of the utmost importance in corroborating such
test. For example, the Doppler tracking data from the Mars
Exploration Rover (MER) to Odyssey will improve the determination
of just the out-of-plane portion of the Odyssey orbit. An
independent check which could, and should, be done is to produce a
particular solution in which it is also introduced a parameter
which explicitly accounts for the Lense-Thirring effect and solve
for it.
\section*{Acknowledgments}
I gratefully thank A. Konopliv, NASA Jet Propulsion Laboratory
(JPL), for having kindly provided me with the entire MGS data set.
I am also grateful to H. Lichtenegger for his kind hospitality at
\"{O}AW Institut f\"{u}r Weltraumforschung (IWF) in Graz.
%
%-----------------------------------------

\end{document}